\documentclass[showpacs,twocolumn,prb,floatfix]{revtex4}
\usepackage{graphicx, times}
\usepackage{amsmath}
\newcommand{\ba}{\begin{eqnarray}}
\newcommand{\be}{\begin{equation}}
\newcommand{\ea}{\end{eqnarray}}
\newcommand{\ee}{\end{equation}}

\begin{document}
\title{Disorder-Induced Superfluidity in Hardcore Bosons in Two Dimensions} 

\author{Ji-Woo Lee}
\altaffiliation{Current address:
Korea Institute for Advanced Study, Cheongnyangni 2-dong, Dongdaemun-gu, Seoul, 130-722, Korea}
\author{Shailesh Chandrasekharan}
\author{Harold U. Baranger}
\affiliation {Department of Physics, Duke University,
 Durham, North Carolina 27708-0305, USA 
}

\date{October 16, 2006}

\begin{abstract}
We study the effect of disorder on hardcore bosons in two dimensions at the SU(2) symmetric ``Heisenberg point''. We obtain our results with quantum Monte Carlo simulations using the directed loop algorithm. In the absence of disorder, the system has no long-range order at finite temperature due to the enhanced symmetry. However, the introduction of a disordered potential, uniformly distributed from $-\Delta$ to $\Delta$, induces a finite-temperature superfluid phase. In particular the diagonal correlation length $\xi$ decreases but the superfluid order-parameter correlation function becomes a power-law. A non-monotonic finite-size behavior is noted and explained as arising due to $\xi$. We provide evidence that at long distances the effects of a weak disordered potential can be mimicked by an effective uniform potential with a root-mean-square value: $\mu_{\rm eff} = \Delta/\sqrt{3}$. For strong disorder, the system becomes a Bose glass insulator.
\end{abstract}


\maketitle

\section{Introduction}

The effect of disorder on two-dimensional strongly correlated systems has been an important problem in condensed matter physics for decades since disorder is inevitable in real systems. Experimentally, ${}^4$He in porous media,\cite{Chan03} Josephson junction arrays,\cite{TakehideJJ06} and thin-film superconductors\cite{Goldman06} are good examples. Bosonic models are appropriate in studying these systems, even though the latter two are microscopically fermionic, because the elementary excitations are bosonic.\cite{fisher}  When the bosons interact with each other via a strong short-range repulsive potential, a hardcore boson Hamiltonian is a good starting point.

The hardcore boson model has been used to study zero-temperature phase transitions in a number of contexts:\cite{BernTroyer02,Scalettar91,Zhang95,ScalettarGKZ95,HebertBSSTD01,BernardetBT02,Cuccoli03,AizenmanLieb04,Rousseau1DSL06,AnandPRL06}  
the simple transition as a function of average filling,\cite{BernTroyer02,HebertBSSTD01,AnandPRL06}
the effect of adding a nearest-neighbor repulsive interaction (competition between superfluidity and checker-board solid),\cite{ScalettarGKZ95,HebertBSSTD01,BernardetBT02}
the transition induced by a staggered potential (also superfluid $\to$ checker-board solid),\cite{AizenmanLieb04,Rousseau1DSL06,AnandPRL06}
and the destruction of superfluidity caused by adding disorder (superfluid $\to$ Bose glass).\cite{Scalettar91,Zhang95,BernardetBT02,AnandPRL06}
One reason for the interest in incorporating nearest-neighbor repulsion is that it allows access to a higher symmetry situation: for a particular value of the repulsion, the Hamiltonian has a full SU(2) symmetry, an enhancement of the usual U(1) symmetry associated with boson number.

A key property at finite temperature is whether the system has a non-zero critical temperature or not. In the clean system, for example, at the SU(2) symmetric point (known as the Heisenberg point in analogy with quantum antiferromagnets), the critical temperature is zero by the Mermin-Wagner theorem.\cite{mermin} Away from this special case, the symmetry of the system is U(1), and one expects that superfluidity is possible at a finite temperature; this has been studied numerically by several groups.\cite{HebertBSSTD01,BernardetBT02,Cuccoli03}

In this paper, we focus on the effect of adding disorder to a system at the SU(2) symmetric point at a finite temperature. Though the mean value of the disordered potential is zero, it does, of course, break the SU(2) symmetry of the system. Thus, we study if disorder can induce superfluidity. 

The Hamiltonian for hardcore bosons with nearest-neighbor repulsion which we study here is
\be
H =  \sum_{ \langle i,j  \rangle } \Big\{ -\frac{1}{2}(b^\dagger_i  b_j + b^\dagger_j b_i )
 + n_i n_j \Big\}
- \sum_i  (\mu_i +2) n_i,
\label{eq:Ham}
\ee
where $b^\dagger_i (b_i)$ is the creation (destruction) operator of a boson at site $i$ on a square lattice of size $L\times L$, $n_i  \!\equiv\! b^{\dagger}_i b_i$ is the boson number operator.  Note that if $\mu_i\!=\!0$ for all sites, the system lies at the Heisenberg point [SU(2) symmetry]. The distribution of the disorder potential, $\mu_i$, is taken to be uniform in the interval $[-\Delta, \Delta]$. We study this model using a quantum Monte Carlo method based on the directed loop algorithm.\cite{sandvik} This algorithm reduces the autocorrelation time so that we may simulate lattices up to size $128 \!\times\! 128$. 

In the weak disorder case, we find strong evidence that superfluidity exists at finite temperature: the disordered potential added upon the Heisenberg point induces finite-temperature superfluidity. However, the system has an interesting finite size effect causing non-monotonic behavior. We understand this effect as arising due to the presence of a large but finite density-density correlation length, which is natural in the presence of a weak uniform potential. When the disorder is strong, the superfluidity eventually disappears; the finite compressibility at this transition shows that the insulating phase is a Bose glass.\cite{fisher}

Our paper is organized as follows. In Sec.~II, we show results for the superfluid order-parameter susceptibility and winding number susceptibility for different disorder strengths, and show that we find finite-temperature superfluidity. Sec.~III discusses the results for different correlation functions and explains the reason behind the interesting finite-size behavior. Results for a uniform potential are presented for comparison in Sec.~IV. Finally, we summarize our results and discuss the nature of the insulating phase in Sec.~V.

\section{Order-parameter susceptibility and Winding number susceptibility}

To quantify whether the system is superfluid or not, we measure two physical observables in our simulations: the order-parameter susceptibility and the winding number susceptibility. The superfluid order-parameter susceptibility is defined as 
\be
\chi_{\rm p} = \frac{T}{4 L^2}\int_0^{1/T}  d\tau
\int_0^{1/T} d{\tau'}
\sum_{{\bf r},{\bf r}'} G_p ({\bf r},\tau,{\bf r'},\tau') 
\ee
where
\be
G_p ({\bf r},\tau,{\bf r'},\tau') =
\langle b({\bf r}, \tau) b^\dagger ({\bf r'}, \tau') 
+  b^\dagger ({\bf r}, \tau) b ({\bf r'}, \tau')
\rangle
\ee
and the winding number susceptibility is
\be
\chi_{\rm w} = \frac{\pi}{2} \Big[
\langle W_x^2 \rangle + 
\langle W_y^2 \rangle  
\Big]
\ee
where $W_{x}$ is the spatial winding of the boson world lines in the $x$ direction, and similarly for $W_{y}$. Here $b ({\bf r}, \tau)$ is the Heisenberg operator defined as 
$ e^{H \tau} b({\bf r}) e^{-H \tau}$, and $\langle \cdots \rangle$ denotes averaging over thermal configurations and disorder realizations. Note that the order-parameter susceptibility is the integral of the two-point off-diagonal correlation function, and recall that the winding number susceptibility is proportional to the superfluid stiffness. 

In the large size limit ($L \!\to\! \infty$), the scaling forms for the susceptibilities are
\be
\chi_{\rm p} \to 
\left\{
\begin{array}{cr}
A L^{2-\eta} & ({\rm superfluid})\\
B  &  ({\rm insulator})
\end{array}
\right.
\label{chip}
\ee
and 
\be
\chi_{\rm w} \to 
\left\{
\begin{array}{cr}
C  & ({\rm superfluid})\\
D \exp(-L/L_0)  &  ({\rm insulator}) \;.
\end{array}
\right.
\label{chiw}
\ee
In the superfluid state, $\eta \le  0.25$ and $C \ge 2$ where the limiting values are obtained right at the Berezinskii-Kosterlitz-Thouless transition.

In all simulations, we fix the temperature at $T \!=\!1/8$ (corresponding to $\beta\!=\!8$) and $\epsilon \!\equiv\! \beta/M \!=\! 1/4$ where $M$ is the the number of Trotter-Suzuki\cite{trotter,suzuki} slices in the imaginary time direction. The lattice size ranges up to $128\!\times\!128$. The total number of loop updates for the thermal average is $10000$, and the minimum number of disorder configurations is 100.

\begin{figure}[t]
\includegraphics[width=6.5cm]{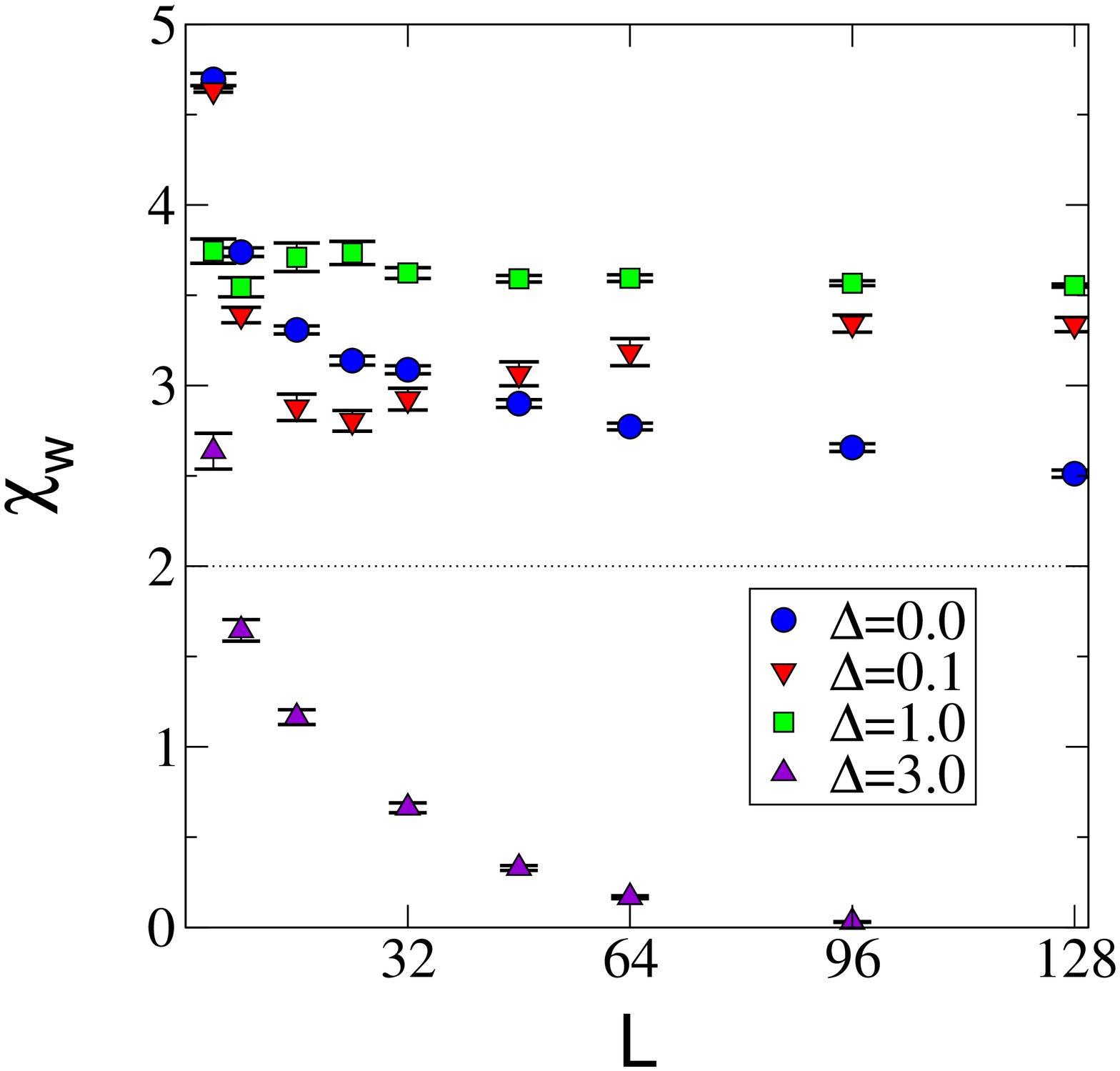}
\includegraphics[width=6.5cm]{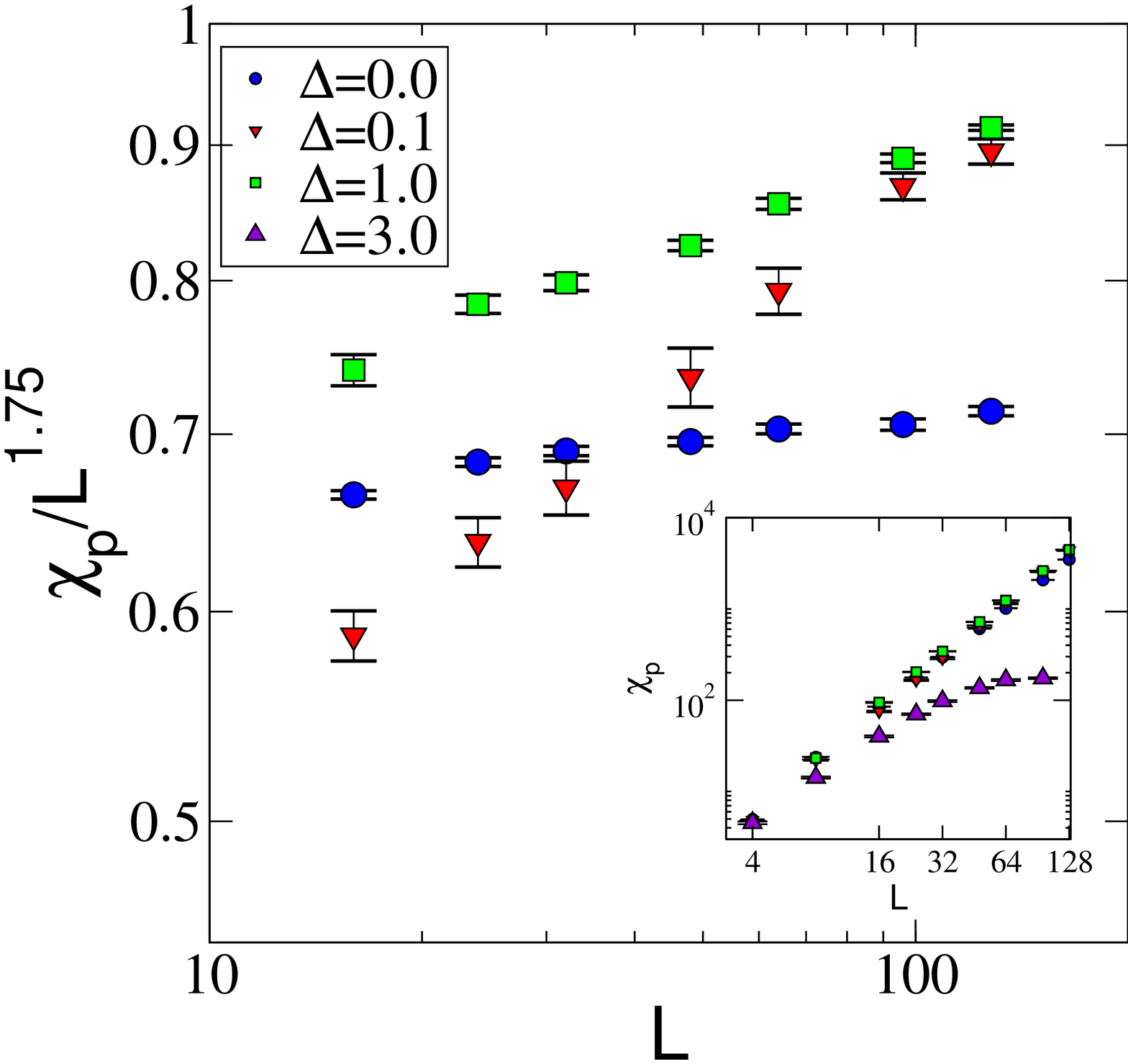}
\vspace*{-0.2in}
\caption{(color online) Winding number susceptibility ($\chi_{\rm w}$, top, linear scale) and order-parameter susceptibility ($\chi_{\rm p}$, bottom, log-log scale) as a function of system size $L$ for different values of the disorder strength $\Delta$. At both $\Delta\!=\!0$ and $\Delta\!=\!3$, the behavior signals the absence of superfluidity. However, at the intermediate values $\Delta\!=\!1.0$ and $0.1$, once the size of the system reaches a certain length scale, both the saturation of $\chi_{\rm w}$ at a value greater than $2$ and the power-law increase of $\chi_{\rm p}$ with exponent greater than $1.75$ signals the presence of superfluidity.
\label{fig1}}
\end{figure}

Figure~\ref{fig1} shows the winding number susceptibility ($\chi_{\rm w}$) and order-parameter susceptibility ($\chi_{\rm p}$) as a function of system size for various disorder strengths. In the clean case, $\Delta\!=\!0$ (Heisenberg point), the winding number susceptibility decreases monotonically as the system size increases. Since there is no ordered phase at finite temperature by the Mermin-Wagner theorem and the correlation length $\xi_0$ is very large\cite{beard},
finite size effects are large. $\chi_{\rm p}$ shows an approximately power-law form; the fit to Eq.\,(4) is given in Table \ref{tb1}. For $\chi_{\rm w}$, we fit to the form $E \!-\! F\log L$ expected from renormalization group arguments;\cite{Azaria} the fit is very good, with $E\!=\!4.37(5)$, $F\!=\!0.38(1)$, and $\chi^2 \!=\!1.2$ over the range $L \!\in\! [16,128]$.

For strong disorder, $\Delta\!=\!3.0$, as the system size increases, the data show that the winding susceptibility decreases exponentially to zero and the order-parameter susceptibility saturates. Thus, as expected, superfluidity disappears in the strongly disordered case. 

\begin{table}[t]
\caption{Fitting results for the order-parameter susceptibility ($\chi_{\rm p}$) and winding number susceptibility ($\chi_{\rm w}$) for $\Delta\!=\!0.0$, $0.1$, and $1.0$. The fitting formulas are Eqs.\,(\ref{chip}) and (\ref{chiw}); the $\chi^2$ given is per degree of freedom from the fit.
}
\begin{tabular}{c||ccc|cc|c}
         & \multicolumn{3}{c|}{$\chi_{\rm p}$} & 
\multicolumn{2}{c|}{$\chi_{\rm w}$} & \\
$\Delta$ &  $A$ & $\eta$ & $\chi^2$  & $C$ & $\chi^2$   &   included $L$  \\
\hline
0.0 & 0.613(5) & 0.218(2) & 2.9 & -  & -  & 16--128 \\
\hline
0.1 & $\approx$0.55 & $\approx$ 0.15 & &   $\approx$ 3.34  &   & 96,128 \\
\hline
1.0 & 0.567(7) & 0.151(3) & 0.9 &  3.572(6)   &  3.2  & 16--128 \\
\end{tabular}
\label{tb1}
\end{table}

The most interesting data is for the intermediate disordered cases, $\Delta\!=\!0.1$ and $1.0$. For $\Delta\!=\!0.1$, the winding number susceptibility first decreases for small lattices but then \textit{increases}, saturating at a large value of nearly 3.4 for sizes larger than $L\!=\!96$. Similarly, the behavior of the order-parameter susceptibility seems to change as a function of $L$. Since we were unable to smoothly fit the data with Eqs.~(\ref{chip}) and (\ref{chiw}), we give a crude value, $\eta \!\approx\! 0.15$, and $C \!\approx\! 3.34$ in Table I obtained from using only two points, $L\!=\!96$ and $L\!=\!128$, by which size both the winding number and order-parameter susceptibilities have become stable. For $\Delta\!=\!1.0$, there is some similar variation for the smallest lattices, but the behavior stabilizes for $L\!>\!16$. Reasonable fits yield $\eta\!=\! 0.151(3)$ and a saturation value of $\chi_w$ of about $3.6$. From the values of $\eta$ and the saturation of $\chi_w$, both of our intermediate disordered cases are superfluid at this temperature.

In summary, we find that there exists a superfluid state when $\Delta\!=\!0.1$ or $1.0$, but that superfluidity is absent at $\Delta\!=\!3.0$ as well as in the clean case. \textit{Thus, weak disorder induces superfluidity which is subsequently destroyed if the disorder becomes too strong.}

\begin{figure}[b]
\includegraphics[width=6.5cm]{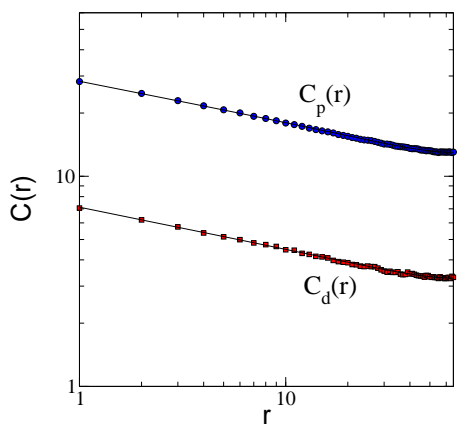}
\vspace*{-0.2in}
\caption{ Order-parameter correlation function, $C_p(r)$, and density-density correlation function, $C_d(r)$, in the clean $\Delta \!=\! 0$ case (log-log scale). The error bar is smaller than the size of the symbols, and the line is a fit to Eq.\,(\ref{fitform}). The $y$-axis is in arbitrary units in order to show that the two correlation functions are the same because of symmetry. We find the exponent $\eta=0.218(2)$ from $C_p(r)$.
\label{fig2}}
\end{figure}

\section{Correlation functions}

The strong and non-monotonic finite size effect in the data of Fig.\,\ref{fig1} remains to be explained. In this Section, we present results for two correlation functions to address this issue. The first is the superfluid order-parameter correlation function defined by
\be
C_p(r) = \frac{T}{L^2}\int_0^{1/T}\!\!  d\tau
\int_0^{1/T}\!\! d{\tau'}
\;{\sum_{{\bf r},{\bf r'}}}' G_p ({\bf r},\tau,{\bf r'},\tau') \;.
\label{cp}
\ee
The second is the density-density correlation function
\be
C_d(r) =  \frac{T}{L^2}\int_0^{1/T}\!\!  d\tau
\int_0^{1/T}\!\! d{\tau'}
\;{\sum_{{\bf r},{\bf r'}}}' G_d ({\bf r},\tau,{\bf r'},\tau')
\label{cd}
\ee
where
\be
G_d({\bf r},\tau,{\bf r'},\tau') = 
\big\langle \sigma({\bf r},{\bf r'}) 
\tilde{n}({\bf r},\tau) \tilde{n}({\bf r}',\tau')  \big\rangle \;.
\ee
with the definition 
$\tilde{n}({\bf r},\tau) \!=\! b^\dagger({\bf r},\tau) b({\bf r},\tau) \!-\! 1/2$ and $\sigma({\bf r},\bf{r'})$ is a stagger factor which is $1$ if ${\bf r}$ and ${\bf r'}$ are on the same sublattice or $-1$ otherwise. We take $r\!\equiv\! x\!-\!x'$, and the sum in Eqs. (\ref{cp}) and (\ref{cd}) is constrained such that $r$ is held fixed and $y \!=\! y'$.

\begin{figure}[b]
\includegraphics[width=6.5cm]{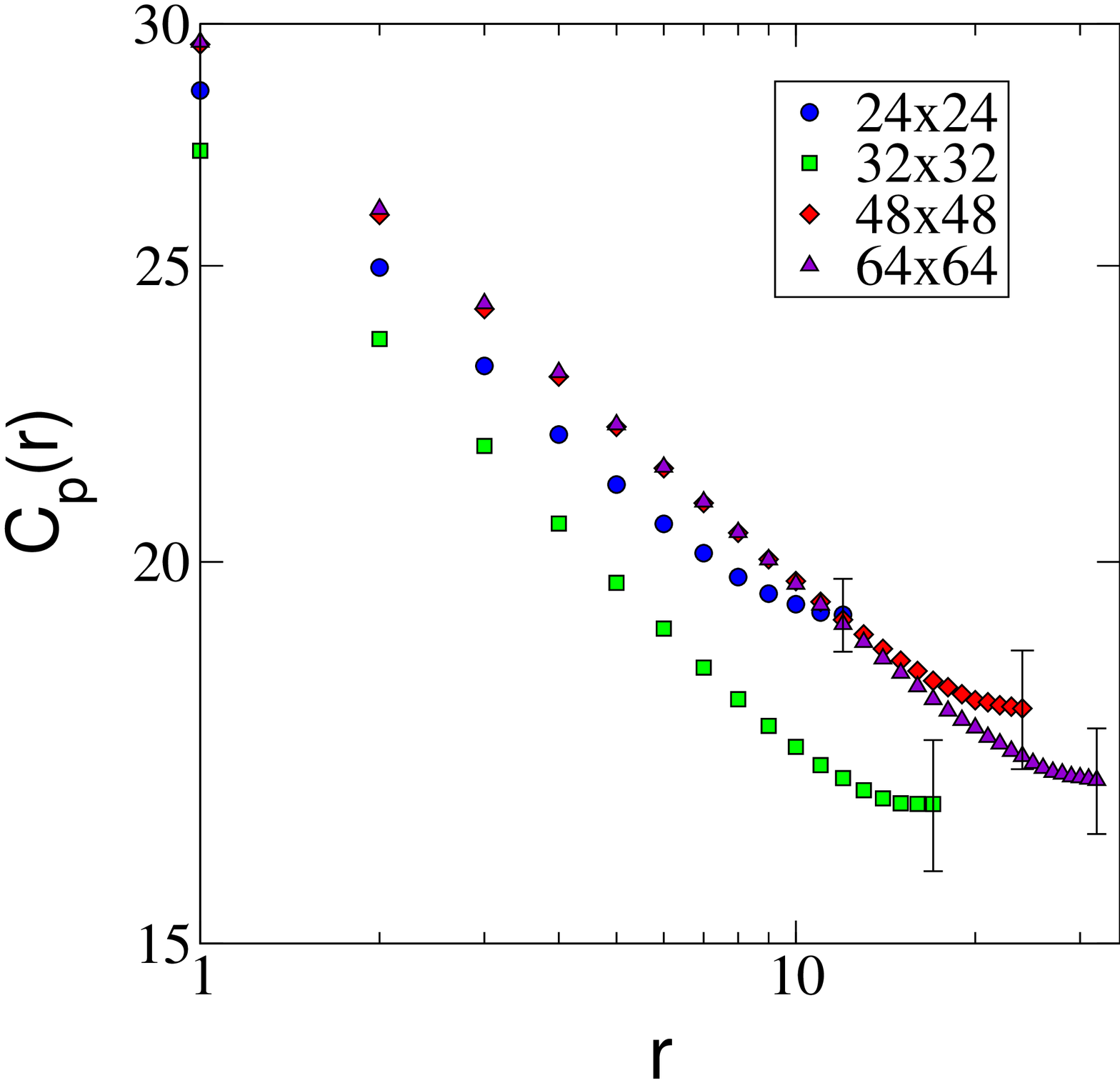}
\includegraphics[width=6.5cm]{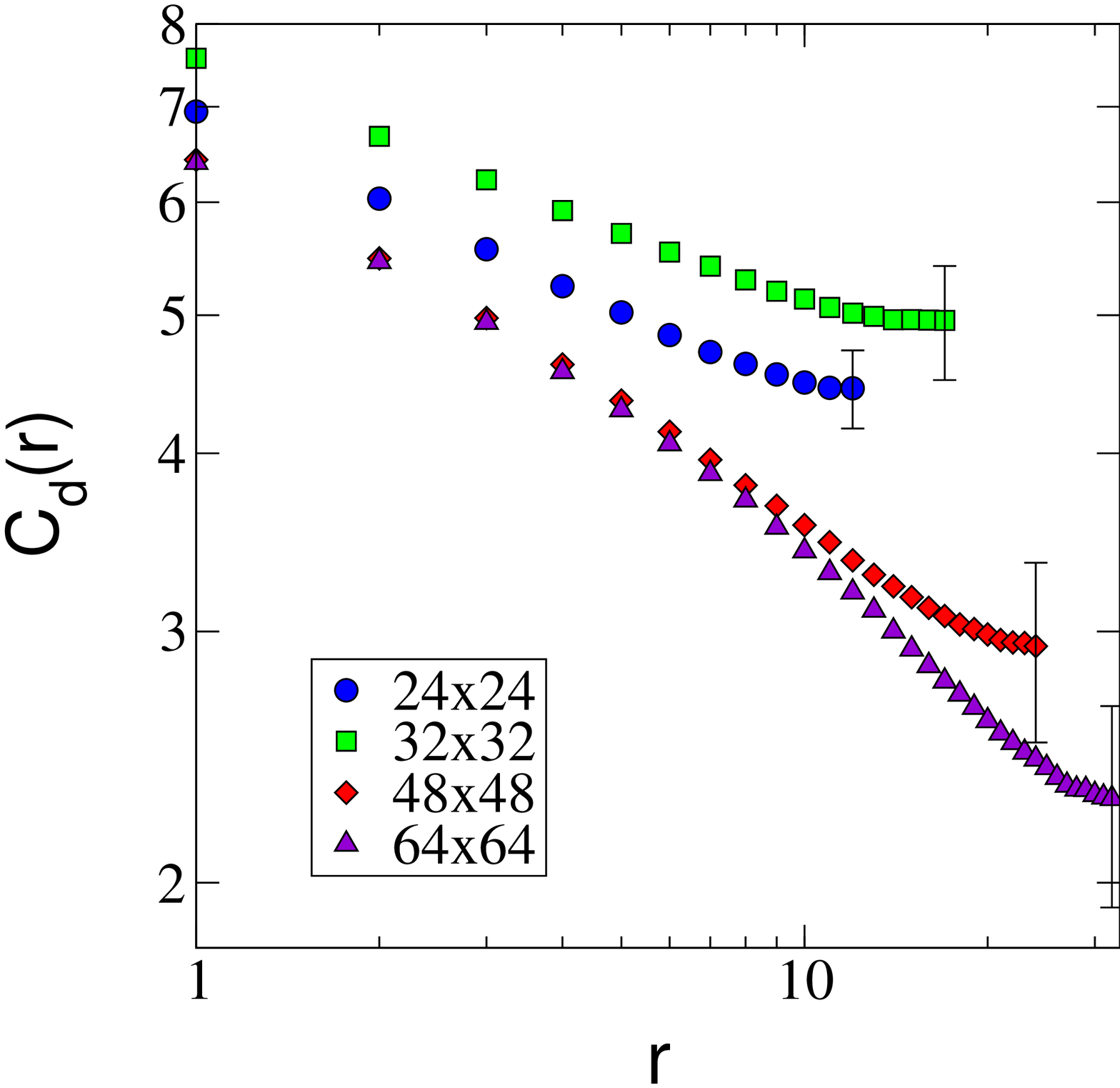}
\vspace*{-0.2in}
\caption{(color online)
Order-parameter correlation function [$C_p(r)$, upper panel] and density-density correlation function [$C_d(r)$, lower panel] for the disordered case $\Delta\!=\!0.1$. $C_p(r)$ has a power-law form for all sizes. In contrast, the behavior of $C_d(r)$ changes from power-law decay to exponential decay for $L \!\ge\! 32$. Thus the diagonal order disappears for larger systems, leaving the system superfluid. We show the error bars of the endpoints.
\label{fig3}}
\end{figure}

In the clean case, $\Delta \!=\! 0$ (Heisenberg point), we expect $C_p(r) \!=\! 4 C_d(r)$ due to the SU(2) symmetry of the model. The two correlation functions are shown in Fig.\,\ref{fig2}. Note that because of the periodic boundary conditions used in the calculation, a correlation function which initially decays must start increasing for $r \!>\! L/2$. Thus, here and throughout this paper, we show the correlation functions only up to $r \!=\! L/2$.

It is well known that when $\Delta \!=\! 0$ the system develops a finite but large correlation length $\xi_0$ at any small temperature. When $L \!\ll\! \xi_0$ it is difficult to detect this correlation length and a power-law behavior should be a good description of the data. Hence we fit our data for $C_p(r)$ to the form $A[1/r^\eta \!+\! 1/(L-r)^\eta]$ and find that $\eta\!=\!0.218(2)$, which is consistent with the fitting of $\chi_{\rm p}$ in Table \ref{tb1}. The small wiggle in $C_d(r)$ arises from the fact that our Monte Carlo update is optimized for measuring the order-parameter susceptibility so that the error in $C_d$ is larger than that in $C_p$.

Now we turn our attention to the disordered case and focus on $\Delta\!=\!0.1$ for which there is a peculiar dip in the winding number susceptibility in Fig.\,\ref{fig1}. First note that in the presence of disorder, the symmetry of the system is affected so that the two correlation functions are no longer identical. In Fig.\,\ref{fig3}, we show both the order-parameter correlation function and the density-density correlation function for $L\!=\!24$, $32$, $48$, and $64$.  Thirty disorder realizations are used for each lattice size. 

Fig.\,\ref{fig3} shows a change in behavior at $L \!\sim\! 30$. For smaller systems, both correlation functions appear power-law. But for larger systems, while $C_p(r)$ is truly power-law, $C_d(r)$ becomes exponential -- a gap develops in the density-density channel. By fitting $C_d(r)$ to the form
\be
C(r) =  A \Big[ e^{-r/\xi} + e^{-(L-r)/\xi} \Big]
\label{fitform}
\ee
we extract $\xi$. We find $\xi_{L\!=\!48} \!=\!17(2)$ and $\xi_{L\!=\!64} \!=\! 19(1)$. These two values suggest that in the thermodynamic limit $\xi$ is roughly $20$.

One can understand the dip in the winding number susceptibility and the bump in the order-parameter susceptibility (Fig.\,\ref{fig1}) with the following scenario: When the system is smaller than $\xi$, the system thinks the full symmetry is still present, and so the winding number decreases as for $\Delta\!=\!0$. But as the system size grows larger than $\xi$, the density-density order is no longer long range, leaving the system in a superfluid state in the thermodynamic limit. As one increases the strength of disorder, the length scale $\xi$ becomes small so that even a small system shows superfluidity, as for $\Delta\!=\!1.0$. Eventually disorder not only destroys the density-density long-range order but also destroys the superfluid long-range order ($\Delta\!=\!3.0$). So we conclude that the weak disorder regime has severe finite size effects but that disorder enhances superfluidity before the system finally enters the strong disorder regime.

\section{Uniform potential}

It is instructive to compare the effect of a disordered potential with that of uniform potential, $\mu_i \!=\! \mu$ for all $i$ in Eq.\,(\ref{eq:Ham}). Since a nonzero uniform potential explicitly breaks the symmetry of the Hamiltonian, it is well known that the system is superfluid at $T\!=\!0$ until the uniform potential reaches a critical value, beyond which the system enters an insulating phase because the hopping is impeded by the the large occupation of particles. The finite-temperature phase diagram was studied in the anisotropic spin-1/2 XXZ model.\cite{BernTroyer02,schmid}  Here we study the isotropic case for comparison with the disordered results above and especially focus on the correlation functions.

\begin{figure}[t]
\includegraphics[width=6.5cm]{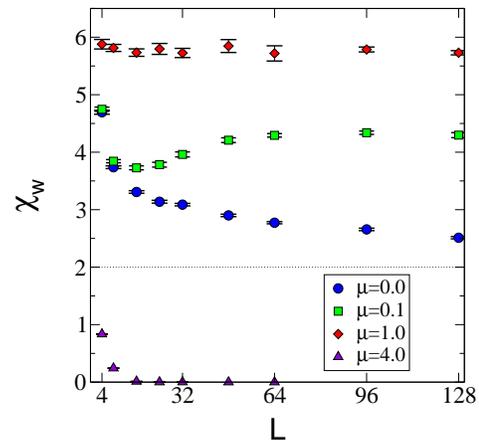}
\vspace*{-0.1in}
\caption{(color online)
For different values of the uniform chemical potential $\mu$, the winding number susceptibility, $\chi_{\rm w}$, as a function of system size. The behavior is very similar to that in the disordered case Fig.\,\ref{fig1}: no superfluidity at either zero or large $\mu$, but substantial superfluidity for intermediate values of $\mu$.
\label{fig4}}
\end{figure}

\begin{figure}[t]
\includegraphics[width=6.5cm]{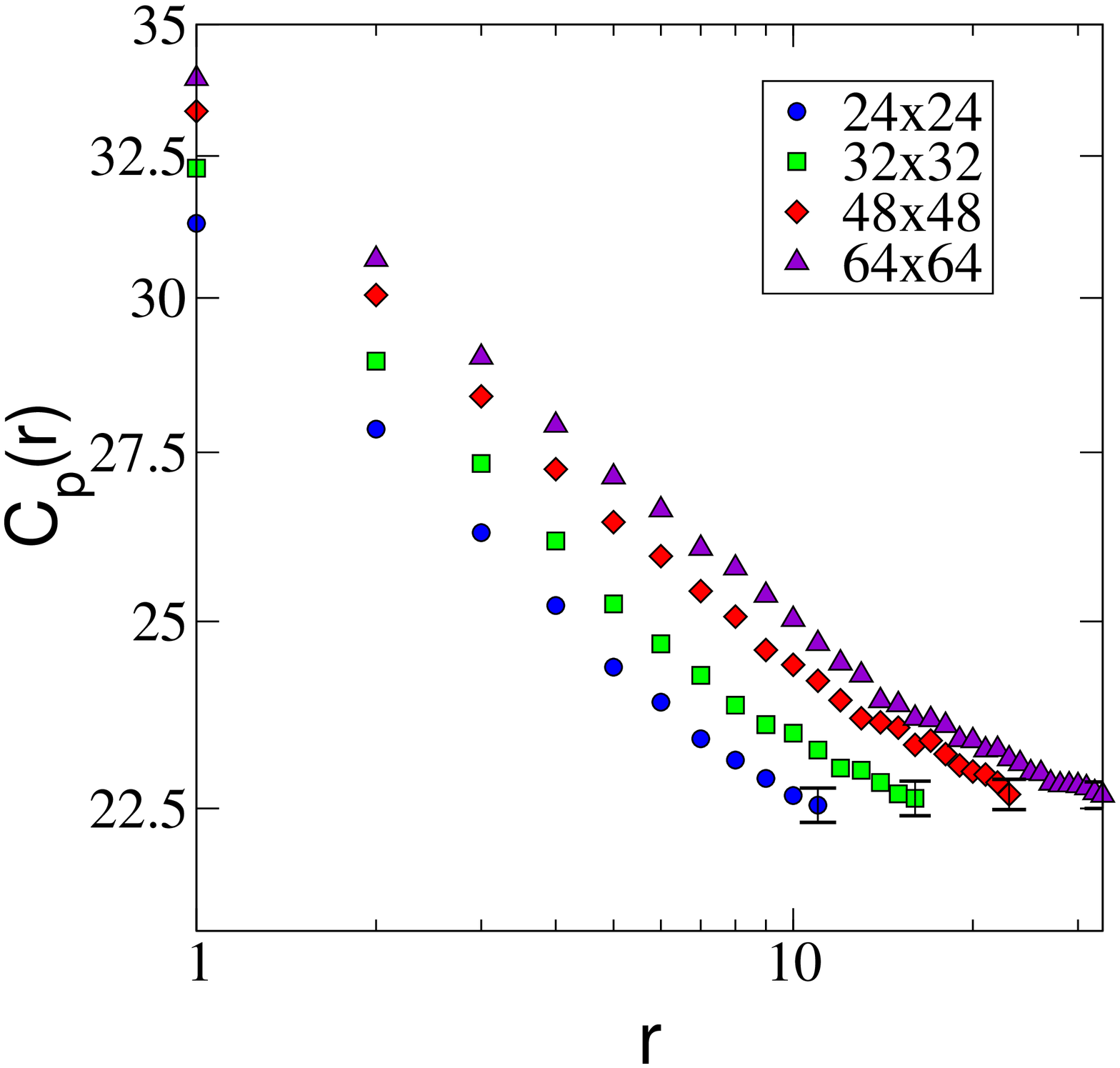}
\includegraphics[width=6.5cm]{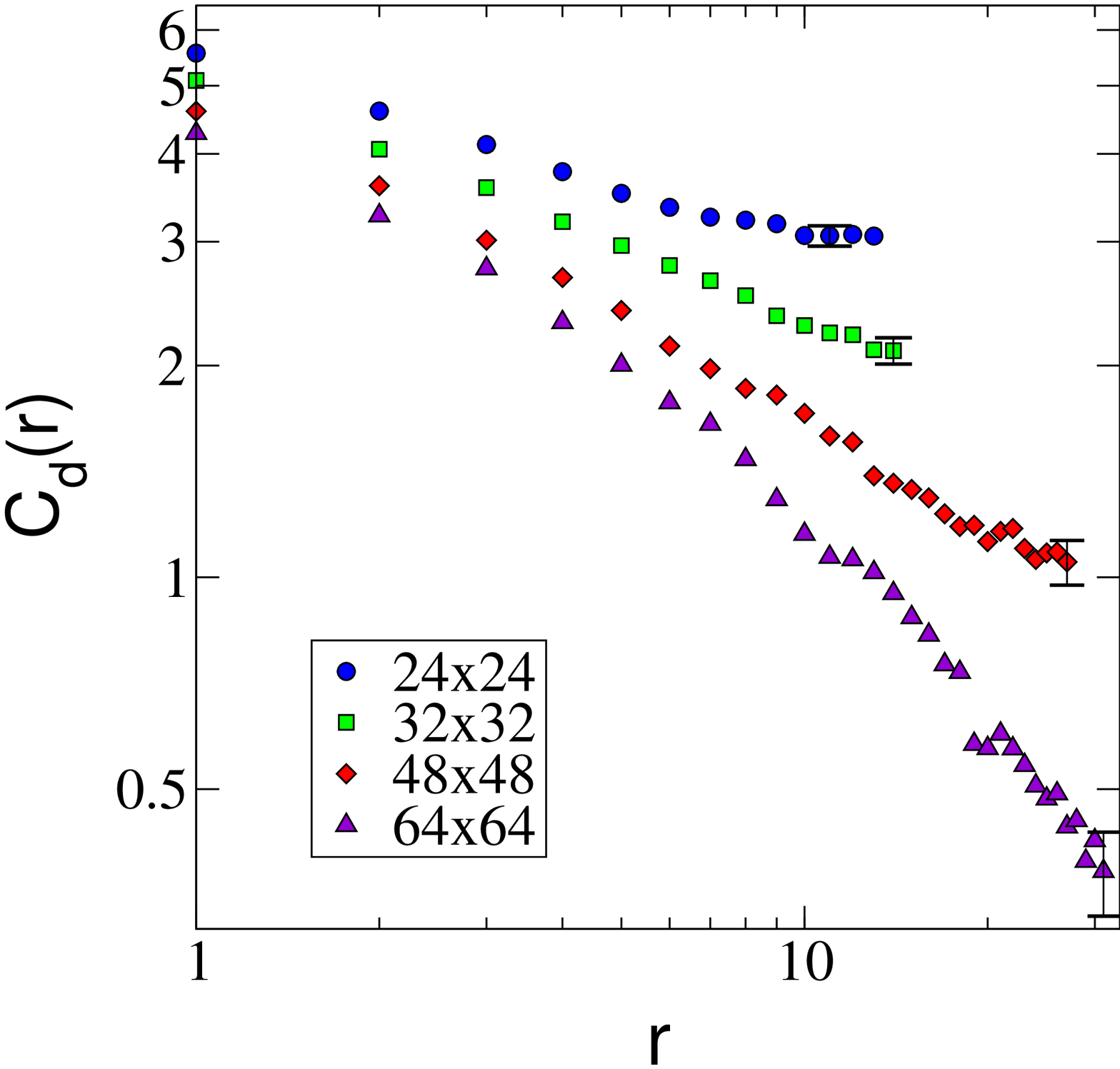}
\vspace*{-0.2in}
\caption{(color online)
For the clean case with uniform potential $\mu\!=\!0.1$, the order-parameter correlation function ($C_p(r)$, upper panel) and density-density correlation function ($C_d(r)$, lower panel). The behavior of $C_p(r)$ is power law, while $C_d(r)$ becomes exponential when it reaches $L=48$. We show the error bars for the endpoints.
\label{fig5}}
\end{figure}

The winding number susceptibility, shown in Fig.\,\ref{fig4}, displays behavior very similar to that in the disordered case. The difference is that even though $\chi_{\rm w}$ decreases when $\mu\!=\!0.1$, the value itself is larger than in the clean ($\Delta\!=\!\mu\!=\!0$) case. We learn that for small sizes weak disordered potential suppresses $\chi_w$ compared to the clean case while a weak uniform potential enhances it. The correlation functions at $\mu\!=\!0.1$ in Fig.\,\ref{fig5} are also similar to those in Fig.\,\ref{fig3}. Using Eq.\,(\ref{fitform}) to fit, we find 
$\xi_{L\!=\!32} \!=\! 13(1)$, $\xi_{L\!=\!48} \!=\! 13.2(5)$, and
$\xi_{L\!=\!64} \!=\! 12.8(5)$. These are all in reasonable accord with each other suggesting that $\xi$ is roughly $13$ in the thermodynamic limit

For uniform potentials, the hardcore boson model given in Eq.\,(\ref{eq:Ham}) is equivalent to a quantum spin-half antiferromagnet in a uniform magnetic field. At zero magnetic field, it is well known that the long wavelength physics of the system is described by a non-linear $O(3)$ sigma model.\cite{Chakravarti1989} In particular an exact formula for the correlation length $\xi_0$ at small temperatures has also been derived using this connection.\cite{Hasenfratz:1990jw} Using these ideas we can write down the effective field theory that must describe the physics of our model for small $\mu$ as long as $\xi_0 \mu \gg 1$. The action of this theory is given by
\be
\frac{\rho}{2T} \int dx\ dy [ 
\partial_x \vec{S} \cdot \partial_x \vec{S}
+
\partial_y \vec{S} \cdot \partial_y \vec{S} - \mu^2  (S_x^2 + S_y^2)
]
\ee
where $\vec{S} \!\equiv\! (S_x,S_y,S_z)$ is a unit three vector. The density-density correlations are mapped into the correlations in $S_z$, while the order parameter correlations are mapped into correlations in $S_x$. Expanding the theory about its minimum then shows that the fluctuations in $S_z$ have a correlation length $\xi \!\propto\! 1/\mu$ while correlations in $S_x$ are long range. Indeed we find that the density-density correlation lengths $\xi \!=\! 12.8(5)$, $5.8(5)$, $4.5(5)$ at $\mu\!=\!0.1$, $0.2$, and $0.3$, respectively (for $L\!=\!64$), agree with this prediction.

\section{Discussion}

\begin{figure}[t]
\includegraphics[width=4.2cm]{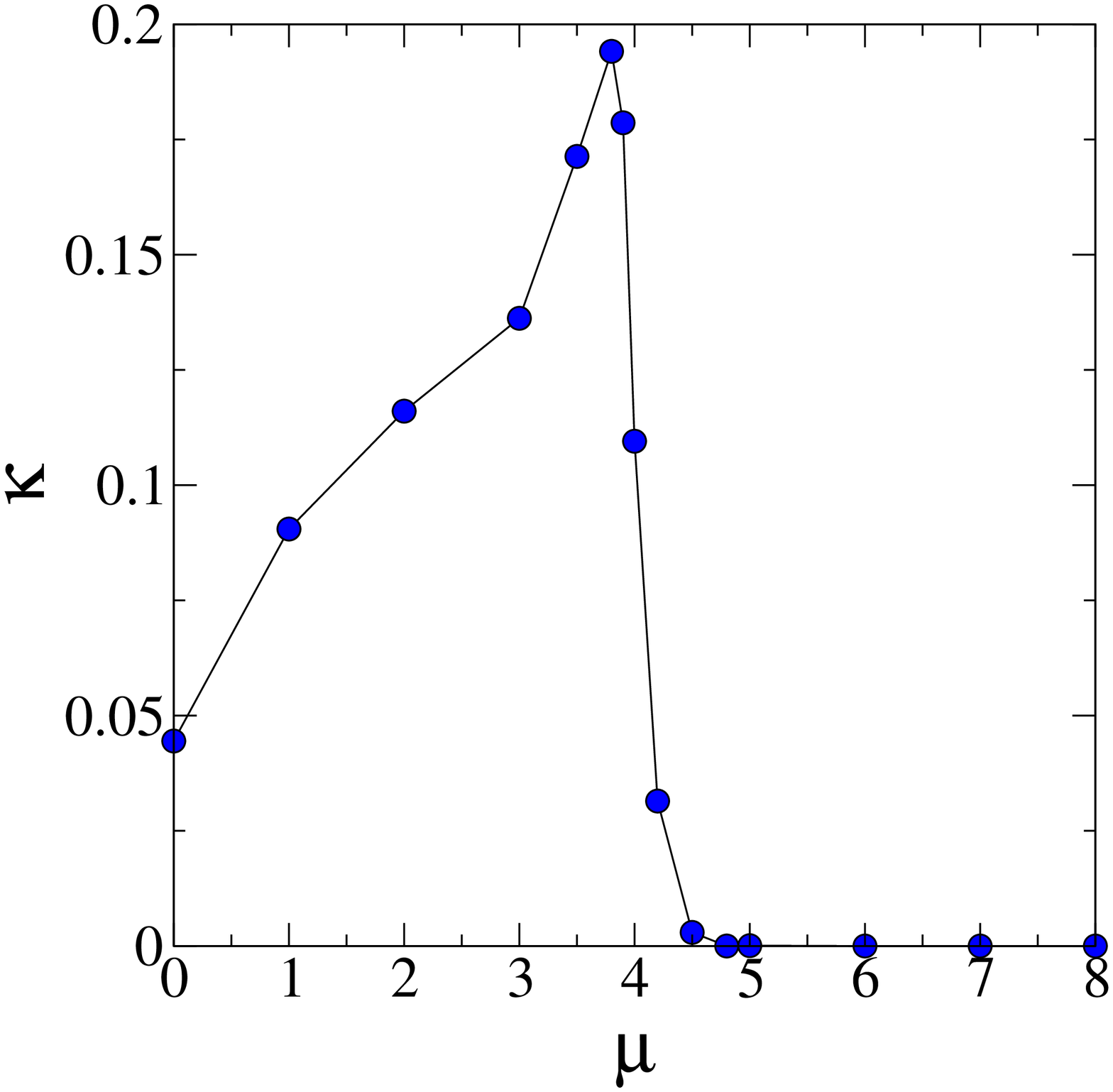}
\includegraphics[width=4.2cm]{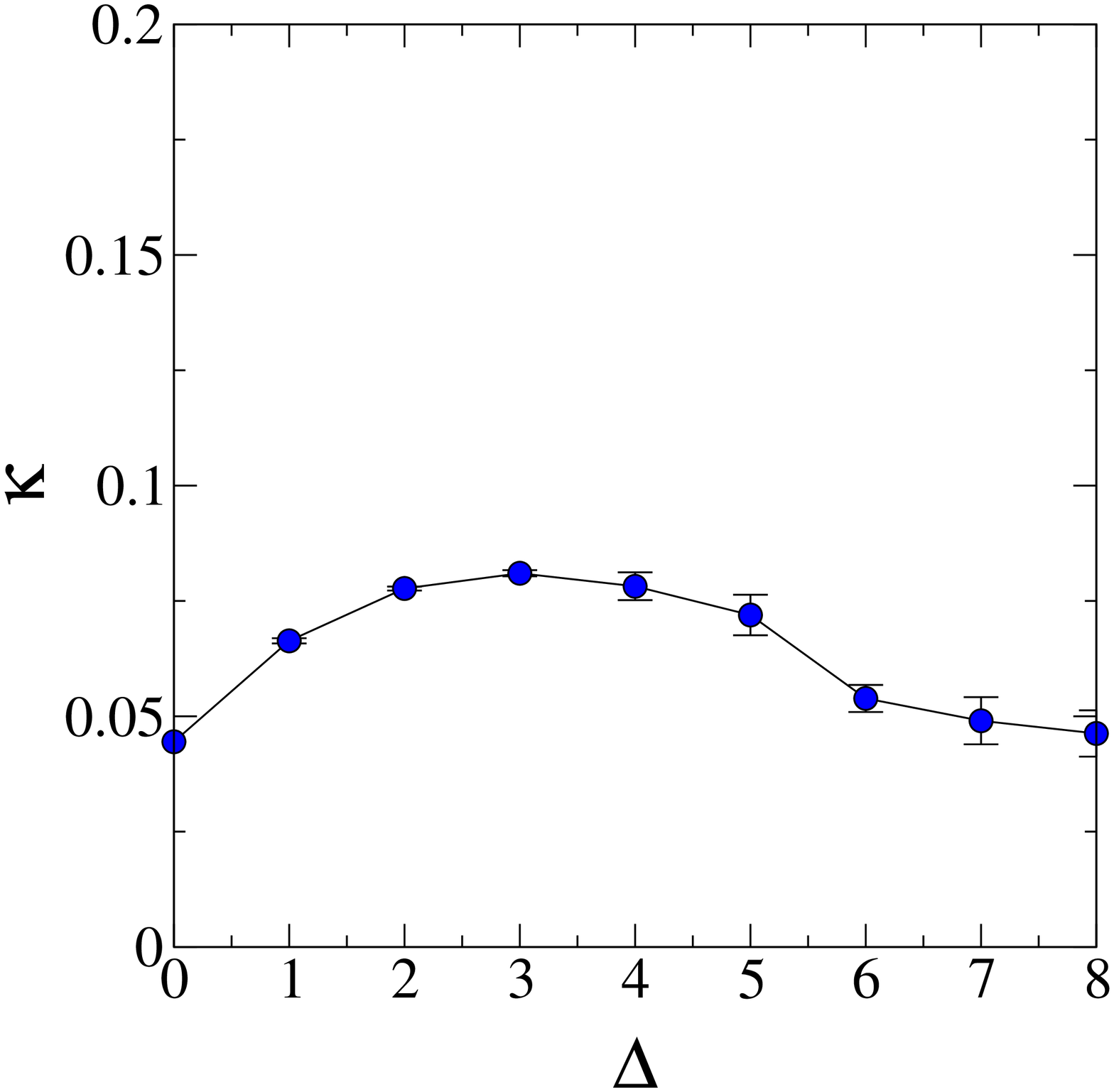}
\vspace*{-0.1in}
\caption{
Compressibility upon applying either a uniform chemical potential (left panel) or disordered potential with zero mean (right panel). The size of the system is $64 \!\times\! 64$, and the number of disorder realizations is 10 in the disordered case. Note that in the disordered case we expect $1 \!<\! \Delta_c \!<\! 3$. Vanishing compressibility indicates that the non-superfluid phase is an incompressible Mott insulator, while finite compressibility indicates a ``Bose glass'' phase.
\label{fig6}}
\end{figure}

In this paper, we have studied the presence or absence of two types of long-range order at finite temperature. When the system has SU(2) symmetry, the superfluid order-parameter correlation function and the density-density correlation function behave identically, and there is no long-range order. We tune away from the SU(2) symmetric point either with a uniform potential or with a disordered potential which has zero mean. In both cases, we show that the density-density correlation function decays exponentially, while the super-fluid order parameter correlation function decays as a power-law, implying that superfluidity exists for sufficiently large systems. For sufficiently large potentials of either type, all correlations again become short range and the superfluidity disappears. 

We can easily understand the existence of superfluidity for small uniform potentials using an effective field theory; it predicts that the density-density correlation length follows $\xi \!\sim\! 1/\mu$.  If we assume that the disordered potential can be replaced by an effective uniform potential equal to the root-mean-square value, we get $\mu_{\rm eff} \!=\! \Delta/\sqrt{3}$. Using our results for the uniform potential, we then predict roughly $\xi \!\sim\! 22(1)$ for $\Delta \!=\! 0.1$ which agrees very well with our results in the disordered case. Thus, we think that a weak disordered potential indeed behaves very much like a uniform chemical potential at long distances.

To understand the phase for large $\mu$ or $\Delta$, we measure the compressibility as a function of $\mu$ or $\Delta$. The compressibility is defined as
\begin{equation}
\kappa \equiv \frac{1}{TL^2}
\Big\langle \Big[\sum_{{\bf r},\tau} n({\bf r},\tau) 
- \Big\langle \sum_{{\bf r},\tau} n({\bf r},\tau) \Big\rangle\Big]^2 \Big\rangle 
\end{equation}
where $n({\bf r},\tau) \!\equiv\!  b^\dag({\bf r}, \tau)b({\bf r}, \tau)$ is the boson number at site $({\bf r}, \tau)$.

In Fig.\,6, we see a striking difference between the compressibility of a clean and disordered system. As the uniform chemical potential $\mu$ increases, the compressibility becomes zero at about $\mu_c \!\sim\! 4$. In contrast, the compressibility in the disordered case remains non-zero, even though we go well beyond $\Delta_c$ which is between $1$ and $3$. The uniform potential increases the boson density until the system becomes an incompressible Mott insulator with density fixed at $1$, since it is hardcore model. But in the disordered case, because the mean potential is $0$, the system remains at half filling, allowing density fluctuations to persist. Nonetheless, the system is no longer superfluid; it enters the ``Bose glass'' phase.\cite{fisher}

We close by emphasizing that finite size effects in this system are more dramatic when disorder is weak: one must average over distances greater than $\xi$ (the density-density correlation length) before the effects of disorder will break the SU(2) symmetry. For smaller system sizes disorder suppresses superfluidity, as measured through $\chi_w$, as compared to the clean case. In the attractive Hubbard-type fermionic model we recently studied,\cite{lee05} with nearest-neighbor repulsive interaction of fermionic pairs, we find similar finite-size behavior of the winding number susceptibility when disorder is weak.\cite{unpublished} If the low-energy theory of this fermionic model is bosonic, as seems to be the case, behavior similar to that described here may occur. For disordered fermion systems, deeper investigation is needed into the true nature of the effects of disorder.

\acknowledgments

The authors thank N. Trivedi and A. Ghosal for helpful discussions. This work was supported in part by National Science Foundation Grant DMR-0506953.

\bibliography{paper,dirtybosons}

\end{document}